\documentclass[conference]{IEEEtran}
\IEEEoverridecommandlockouts
\usepackage{cite}
\usepackage{amsmath,amssymb,amsfonts}
\usepackage{algorithmic}
\usepackage{graphicx}
\usepackage{textcomp}
\usepackage{xcolor}

\usepackage{booktabs} 
\usepackage{adjustbox}
\usepackage{multirow}
\usepackage{enumitem}

\newcommand{\hl}[1]{#1}

\usepackage[linesnumbered,algoruled,boxed,lined]{algorithm2e}
\usepackage{amsthm}
\usepackage{pgfplots}
\usepackage{filecontents}
\usepackage{mathtools}
\usetikzlibrary{shapes,snakes,positioning}

\definecolor{tblue}{RGB}{31,119,180}
\definecolor{torange}{RGB}{255,127,14}
\definecolor{tgreen}{RGB}{44,160,44}
\definecolor{tred}{RGB}{214,39,40}
\definecolor{tpurple}{RGB}{148,103,189}

\newcommand{\eg}{e.g.}
\newcommand{\ie}{i.e.}
\newcommand{\nop}[1]{}
\theoremstyle{definition}
\newtheorem{mydef}{Definition}
\newcommand{\argmax}{\arg\,\max} 
\newcommand{\ignore}[1]{}

\def\BibTeX{{\rm B\kern-.05em{\sc i\kern-.025em b}\kern-.08em
    T\kern-.1667em\lower.7ex\hbox{E}\kern-.125emX}}
\begin{document}

\title{Representation Learning in Heterogeneous Professional Social Networks with Ambiguous Social Connections}

\author{\IEEEauthorblockN{Baoxu Shi}
\IEEEauthorblockA{\textit{LinkedIn} \\
dashi@linkedin.com}
\and
\IEEEauthorblockN{Jaewon Yang}
\IEEEauthorblockA{\textit{LinkedIn} \\
jeyang@linkedin.com}
\and
\IEEEauthorblockN{Tim Weninger}
\IEEEauthorblockA{\textit{University of Notre Dame} \\
tweninge@nd.edu}
\and
\IEEEauthorblockN{Jing How}
\IEEEauthorblockA{\textit{Pinterest} \\
kublai.jing@gmail.com}
\and
\IEEEauthorblockN{Qi He}
\IEEEauthorblockA{\textit{LinkedIn} \\
qhe@linkedin.com}
}

\maketitle

\begin{abstract}
Network representations have been shown to improve performance within a variety of tasks, including classification, clustering, and link prediction. However, most models either focus on moderate-sized, homogeneous networks or require a significant amount of auxiliary input to be provided by the user. Moreover, few works have studied network representations in real-world heterogeneous social networks with ambiguous social connections and are often incomplete. In the present work, we investigate the problem of learning low-dimensional node representations in heterogeneous professional social networks (HPSNs), which are incomplete and have ambiguous social connections. We present a general heterogeneous network representation learning model called Star2Vec that learns entity and person embeddings jointly using a social connection strength-aware biased random walk combined with a node-structure expansion function. Experiments on LinkedIn's Economic Graph and publicly available snapshots of Facebook's network show that Star2Vec outperforms existing methods on members' industry and social circle classification, skill and title clustering, and member-entity link predictions. We also conducted large-scale case studies to demonstrate practical applications of the Star2Vec embeddings trained on LinkedIn's Economic Graph such as next career move, alternative career suggestions, and general entity similarity searches.
\end{abstract}


\section{Introduction}\label{sec:introduction}

Many important tasks in network analysis, for example node classification~\cite{bhagat2011node}, community detection~\cite{fortunato2010community}, recommendation~\cite{zhou2007bipartite}, and link prediction~\cite{liben2007link}, rely primarily on the discovery and modelling of patterns hidden among the nodes and edges in a network. To that end, recent advances in network-based Representation Learning (RL) have shown the ability to capture these patterns within a vector-embedding of the network's nodes. These embeddings can then be used for a variety of network analysis tasks. 

\hl{In order to learn good network embeddings on very large heterogeneous professional social networks (HPSNs) such as LinkedIn's Economic Graph and power entity recommendation or entity retrieval based products, an HPSN representation learning model needs to:}

\begin{itemize}[leftmargin=*]
    \item \hl{utilize the rich type information in HPSNs,}
    \item \hl{learn both person and entity representations in the same semantic space,}
    \item \hl{handle ambiguous and unobserved social connections in HPSNs,}
    \item \hl{handle networks with hundreds-of-millions of nodes and billions of edges.}
\end{itemize}

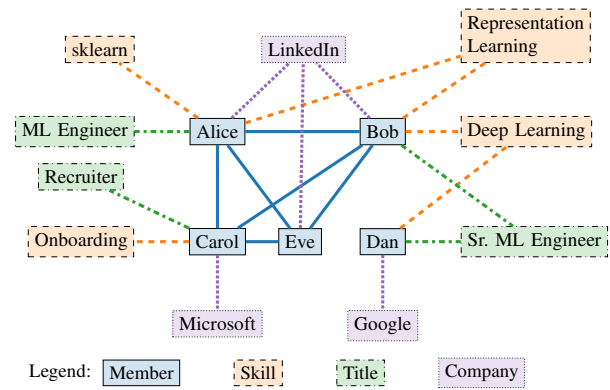
\begin{figure}[t]
\centering
\begin{adjustbox}{max width=.90\linewidth}
    \begin{tikzpicture}
\tikzstyle{arrow} = [ultra thick]
\tikzstyle{line} = [ultra thick]
\tikzstyle{member} = [fill=tblue!20]
\tikzstyle{skill} = [fill=torange!20,dashed]
\tikzstyle{title} = [fill=tgreen!20,dashdotted]
\tikzstyle{company} = [fill=tpurple!20,densely dotted]
\node[draw, member, name=Alice] at (0,0) {Alice};
\node[draw, member, name=Bob] at (3, 0) {Bob};
\node[draw, member, name=Carol] at (0, -2) {Carol};
\node[draw, member, name=Dan] at (3, -2) {Dan};
\node[draw, member, name=Eve] at (1.5, -2) {Eve};

\node[draw, skill, name=sklearn, above left = of Alice] {sklearn};
\node[draw, skill, name=deeplearning, right = of Bob] {Deep Learning};
\node[draw, skill, name=onboarding, left = of Carol] {Onboarding};
\node[draw, skill, name=replearning, align=left, above right = of Bob] {Representation\\Learning};

\node[draw, title, name=mlengineer, left = of Alice] {ML Engineer};
\node[draw, title, name=recruiter, above = .65 of onboarding] {Recruiter};
\node[draw, title, name=srmlengineer, right = of Dan] {Sr. ML Engineer};

\node[draw, company, name=msft, below = of Carol] {Microsoft};
\node[draw, company, name=lkdn, above = of Eve, right = 2.3 of sklearn] {LinkedIn};
\node[draw, company, name=goog, below = of Dan] {Google};

\draw [line,tblue] (Alice) -- (Bob);
\draw [line,tblue] (Alice) -- (Carol);
\draw [line,tblue] (Bob) -- (Carol);
\draw [line,tblue] (Alice) -- (Eve);
\draw [line,tblue] (Bob) -- (Eve);
\draw [line,tblue] (Carol) -- (Eve);

\draw [arrow,dashed,torange] (Alice) -- (sklearn);
\draw [arrow,dashed,torange] (Alice) -- (replearning);
\draw [arrow,dashed,torange] (Bob) -- (deeplearning);
\draw [arrow,dashed,torange] (Bob) -- (replearning);
\draw [arrow,dashed,torange] (Carol) -- (onboarding);
\draw [arrow,dashed,torange] (Dan) -- (deeplearning);

\draw [arrow,dashdotted,tgreen] (Alice) -- (mlengineer);
\draw [arrow,dashdotted,tgreen] (Carol) -- (recruiter);
\draw [arrow,dashdotted,tgreen] (Bob) -- (srmlengineer);
\draw [arrow,dashdotted,tgreen] (Dan) -- (srmlengineer);

\draw [arrow,densely dotted,tpurple] (Alice) -- (lkdn);
\draw [arrow,densely dotted,tpurple] (Bob) -- (lkdn);
\draw [arrow,densely dotted,tpurple] (Carol) -- (msft);
\draw [arrow,densely dotted,tpurple] (Dan) -- (goog);
\draw [arrow,densely dotted,tpurple] (Eve) -- (lkdn);

\end{tikzpicture}
\end{adjustbox}

\vspace{2mm}\begin{adjustbox}{max width=.65\linewidth}
    \begin{tikzpicture}[trim left=-8mm]
\tikzstyle{member} = [fill=tblue!20]
\tikzstyle{skill} = [fill=torange!20,dashed]
\tikzstyle{title} = [fill=tgreen!20,dashdotted]
\tikzstyle{company} = [fill=tpurple!20,densely dotted]

\node[draw, member, name=mem] at (0,0) {Member};
\node[left = .1 of mem] {Legend:};
\node[draw, skill, name=sk, right = of mem] {Skill};
\node[draw, title, name=ti, right = of sk] {Title};
\node[draw, company, name=cmp, , right = of ti] {Company};
\end{tikzpicture}
\end{adjustbox}
\caption{Example professional social network represented as a heterogeneous information network.}
\label{fig:graph}
\vspace{-0.3cm}
\end{figure}

\hl{Unfortunately, few systems have addressed all four requirements.} Most existing methods in this area focus on homogeneous (\ie, untyped) networks~\cite{perozzi2014deepwalk,tang2015line,grover2016node2vec}, which assumes that all nodes share a single node type. Although these shallow models have fewer parameters and can run on very large networks, they ignore heterogeneity found in many networks. Recently there has been some work to extend homogeneous network models through the use of auxiliary features like node attribute or content~\cite{huang2017label,chang2015,liao2018attributed,gao2018deep}. However, using supplemental features explodes the parameter space and is prone to overfitting.

Although these augmented models achieve promising results on homogeneous networks, \hl{real-world social networks, such as LinkedIn and Facebook, are often heterogeneous networks with multiple node types like person, school, company, interest, and many others. Therefore, one major} challenge with learning representations in \hl{heterogeneous social networks} is to find proper ways to leverage their rich type information. Metapath2vec~\cite{dong2017metapath2vec} introduced a metapath-based method that learns node-embeddings from heterogeneous networks; but this approach is limited by human-curated metapaths, which requires domain-specific knowledge and can be difficult to generate on heterogeneous social networks with complex semantics or a large number of node types. \hl{On the other hand, researchers have also tried to model heterogeneous social networks as attributed graphs where persons are the nodes in the network and all other non-person entities are treated as attributes. These methods such as SNE~\cite{liao2017sne} and LANE~\cite{huang2017label} treat attribute entities as input features instead of nodes and cannot learn entity and person embeddings jointly.}

\hl{Besides utilizing rich type information and learning both entity and person embeddings, an HPSN representation learning model also needs to handle both unobserved and observed but ambiguous social connections that exist in professional heterogeneous social networks.} The majority of social networks are incomplete and do not contain all the social connections people have in the real world. Moreover, the social connections captured in social networks are often ambiguous. In real-world scenarios, relationships often take many forms; for example, a person-to-person social connection in an HPSNs could represent spouse, coworker, acquaintance, etc~\cite{henderson2012rolx}. However, most networks do not distinguish among these relationships and often use a single semantically ambiguous relationship \textit{connectedTo} or a few coarse relationships,~\eg, \textit{friend} and \textit{follow}, to represent them all~\cite{liben2007link}. 

\hl{Social networks usually model social connections with a limited number of ambiguous relationship types because} it is often unfeasible to \hl{automatically or even} manually disambiguate these relationships, not only because such a task is costly but also because the relationship's granularity and interpretation are subjective. Throughout the present work we will reference a simplified example network from LinkedIn illustrated in Fig.~\ref{fig:graph} to aid in our discussion. In Fig.~\ref{fig:graph}, the social connection between \texttt{Alice} and \texttt{Carol} can be either candidate-recruiter relationship or college friends. \texttt{Alice} and \texttt{Bob}, on the other hand, can be either close coworkers or acquaintances who work at the same company. Moreover, the social connection between people are transient and evolves over time, which makes it even harder to disambiguate social connections. Again take Fig.~\ref{fig:graph} as an example, \texttt{Carol} and \texttt{Eve} might be coworkers at some point and later becomes recruiter-candidate relationship after Eve moves to LinkedIn. 
Ideally, an HPSN representation learning model ought to be able to infer unobserved social connections and distinguish between different ambiguous social connection types and their social strengths. Unfortunately, most existing network embedding models ignore the network incompleteness and the edge heterogeneity, and simply assume social networks are complete and all existing social connections are the same~\cite{grover2016node2vec}. \hl{As a result, those models will not handle persons with limited number of social connections well and learn similar representations for nodes with common social connections regardless the actual social relationship types and connection strength.}

In addition to direct connections, ambiguous social connections also affect higher-order proximity which is also used to measure node similarity in representation learning models~\cite{tang2015line}. An implicit assumption used in these models is that indirect (\ie, second- or third-order) connections between two nodes are reliable, but this assumption does not apply when there are \hl{ambiguous social connections with different social strengths.} In fact, because direct social connections do \hl{not always represent the same degree of similarity}, the error will cascade and cause further problems when using higher-order proximity. Consider again Fig.~\ref{fig:graph}, wherein second-order proximity models that count common neighbors of, say, \texttt{Alice} and \texttt{Carol} will be misled by \texttt{Eve} and \texttt{Bob} to believe that \texttt{Alice} and \texttt{Carol} have similar characteristics. 

\hl{The presence of ambiguous social connections affects most  network representation models. Higher-order proximity models~\cite{tang2015line,wang2016structural,yang2017fast} overlook this issue and implicitly assume that network connections are always reliable in all scenarios. This problem also has a more severe impact on all random walk-based models~\cite{hamilton2017representation} because these models optimize node representations based on a false assumption that nodes connected within $k$-steps are similar.} In response, more recent network models~\cite{pan2016tri,huang2017label} include text and labels as additional signals. Although they show promising results by adding more parameters, they do not get to the root of the ambiguous social connection problems discussed above.

In the present work, we develop a fast and scalable HPSN representation learning model called Star2Vec that requires little human supervision, \hl{ contains no auxiliary features, and can run on very large real-world networks.
To address the problems raised above, Star2Vec has the ability to 1) automatically weight social connections and leverage unobserved social connections based on heterogeneous second-order proximity, and 2) learn person and non-person entity embeddings jointly with a node structure expansion mechanism}. In summary we make the following contributions:




\begin{itemize}[leftmargin=*]
\item We describe Star2Vec, a scalable model that learns person and entity representations on very large heterogeneous professional social networks,
\item \hl{We introduce a social connection strength-aware random walk model to overcome social connection ambiguity and leverage unobserved social connections without increasing the number of model parameters},
\item \hl{We introduce a node-structure expansion model to expand person node into person-entity structures and learn person and non-person entity embeddings in the same space,}
\item \hl{We perform extensive experiments on LinkedIn's Economic Graph and show the effectiveness of the learned HPSN representations in a variety of tasks,}
\item \hl{We demonstrate that Star2Vec can be applied to other heterogeneous social networks by evaluating Star2Vec on available portions of Facebook.}
\end{itemize}

\hl{The following of the work is organized as follows. We first give formal definitions of the network representation learning task on HPSNs in Sec.~\ref{sec:problem_definition}. Section~\ref{sec:star2vec} gives a detailed description of the proposed Star2Vec model. Then we present our evaluation results on LinkedIn and Facebook datasets in Sec.~\ref{sec:experiments}, followed by a discussion of related works in Sec.~\ref{sec:related_work}.}

\section{Problem Definition}\label{sec:problem_definition}

In this work, we define a heterogeneous professional social network (HPSN) as a network where both nodes and edges are labeled~\cite{shi2017hinsurvey}. The formal definition of HPSN is as follows

\begin{mydef}\label{def:HSN}
A \hl{\textbf{Heterogeneous Professional Social Network}} is a graph $\mathcal{G}=(\mathbf{V},\mathbf{E},\mathbf{T},\mathbf{R}$) in which $\mathbf{V}$, $\mathbf{E}$, $\mathbf{T}$, $\mathbf{R}$ are nodes, edges, node types, and edge types respectively. \hl{$\texttt{person}\in\mathbf{T}$ and $\mathbf{R}=\mathbf{R}_{ppl}\cup\mathbf{R}_{ent}$, where $\mathbf{R}_{ppl}$ are person-to-person edge types and $\mathbf{R}_{ent}$ are person-to-entity edge types.} $|\mathbf{T}| \geq 2$, $|\mathbf{R}_{ppl}| \geq 1$, and $|\mathbf{R}_{ent}| \geq 1$. Each node $u$ is associated with a type mapping function $\phi(\cdot)$ defined as $\phi(u) = T_u, T_u \in \mathbf{T}$. Similarly, each edge $e = u\rightarrow v$ is associated with a relationship type mapping function $\psi(e) = r, r \in \mathbf{R}$.
\end{mydef}

\hl{Definition~\ref{def:HSN} defines an HPSN as a social network having more than one type of node and more than one type of edge. The main difference between an HPSN and other definitions of heterogeneous social networks is that instead of modeling casual social connections, an HPSN focuses on professional entities and emphasizes professional social relationships. For example, LinkedIn's HPSN, call the Economic Graph, contains person-nodes as well as entities such as skills, titles, schools, degrees, companies, jobs, and many other professional entities. As for edge types, besides the main person-to-person social connections, there also exist professional connections such as person-worksAt-company, person-knows-skill, etc.}



Next, we define the path-based network representation learning task on HPSNs to contain two sub-tasks: path generation and \hl{path-based} network representation learning.

\begin{mydef}\label{def:pc}
Given an HPSN $\mathcal{G}$, \textbf{Path Generation} constructs a collection of paths $\mathbf{P} = \{u_0 \leadsto u_l\}$ as the input to the network representation learner, where $u_0 \leadsto u_l$ denotes some length-$l$ path. The path generation process extends a length-$i$ path $u_0\leadsto u_i$ to a length-$(i+1)$ path $u_0\leadsto u_i \rightarrow u_{i+1}$ based on a \hl{biased random walk transition scoring function $\mathcal{P}(u_{i}, u_{i+1},\mathcal{G})$, which determines the probability of walking from $u_i$ to $u_{i+1}$ on graph $\mathcal{G}$ with respect to the social connection strength between $u_i$ and $u_{i+1}$}.
\end{mydef}

\begin{mydef}\label{def:nrl}
Given a collection of paths $\mathbf{P}$ and an HPSN $\mathcal{G}$, \textbf{Path-based Network Representation Learning} learns a $\mathbf{W} \in \mathbb{R}^{|\mathbf{V}|\times d}$ node embedding matrix in which $d \ll |\mathbf{V}|$ that minimizes some loss function $\mathcal{L}(\cdot)$ using path set $\mathbf{P}^\prime = \{\mathcal{F}(p, \mathcal{G})|p\in \mathbf{P}\}$ s.t. $\mathcal{F}(\cdot)$ is some optional post-processing function.
\end{mydef}

Recall the primary challenge in learning embeddings on HPSNs is generating meaningful paths that \hl{carry reliable semantic meaning\cite{dong2017metapath2vec}}. 
\hl{However, in most real-world social networks, generating such paths becomes more challenging due to the connection ambiguity of its social connections.}
\hl{As we state before, heterogeneous social networks like LinkedIn and Facebook primarily use a coarse person-to-person relationship type to denote a variety of social connection types including but not limited to coworkers, friends, acquaintance, and many others. When learning person and entity representations on those networks, a model should be able to properly weight such ambiguous person-to-person social connections so that the ones carry less relevance signals,~\ie, social strength, will play a less important role compared to other social connections.}

Therefore, the main focus here is how to design a biased transition function $\mathcal{P}(\cdot)$ \hl{that properly weights ambiguous social connections in HPSN. So we design the corresponding post-processing function $\mathcal{F}(\cdot)$ to generate high quality paths, and the loss function $\mathcal{L}(\cdot)$ to train the person and non-person entity representations accordingly.}

\section{Star2Vec}\label{sec:star2vec}

\begin{figure*}[ht]
\centering
\begin{adjustbox}{max width=.8\linewidth}
\begin{tikzpicture}
\tikzstyle{arrow} = [ultra thick,->,>=stealth]
\tikzstyle{line} = [thick]
\tikzstyle{member} = [fill=tblue!20]
\tikzstyle{skill} = [fill=torange!20]
\tikzstyle{title} = [fill=tgreen!20]
\tikzstyle{company} = [fill=tpurple!20]

\node[draw, member, name=Alice_1, opacity=0.4] at (0,0) {Alice};
\node[draw, member, name=Eve_1, right = of Alice_1, opacity=0.4] {Eve};
\node[draw, member, name=Carol_1, right = of Eve_1] {Carol};
\node[draw, skill, name=Onboarding, right = of Carol_1] {Onboarding};
\node[draw, member, name=Carol_2, right = of Onboarding, ultra thick] {Carol};
\node[draw, member, name=Bob_1, right = of Carol_2] {Bob};
\node[draw, member, name=Alice_2, right = of Bob_1] {Alice};
\node[draw, skill, name=RepLearning, right = of Alice_2, opacity=0.4] {Rep. Learning};
\node[draw, member, name=Bob_2, right = of RepLearning, opacity=0.4] {Bob};

\draw [arrow] (Alice_1) -- (Eve_1);
\draw [arrow] (Eve_1) -- (Carol_1);
\draw [arrow] (Carol_1) -- (Onboarding);
\draw [arrow] (Onboarding) -- (Carol_2);
\draw [arrow] (Carol_2) -- (Bob_1);
\draw [arrow] (Bob_1) -- (Alice_2);
\draw [arrow] (Alice_2) -- (RepLearning);
\draw [arrow] (RepLearning) -- (Bob_2);

\node[left = of Alice_1] {Traditional path-based skip gram model:};

\path[draw]
    (Carol_2.north) edge[->,>=stealth, bend right=45, dashed] node[below] {$\textsf{Pr}(c_1|u)$} (Onboarding.north)
    (Carol_2.north) edge[->,>=stealth, bend left=45, dashed] node[below] {$\textsf{Pr}(c_2|u)$} (Bob_1.north)
    (Carol_2.south) edge[->,>=stealth, bend left=15, dashed] node[below] {$\textsf{Pr}(c_3|u)$} (Carol_1.south)
    (Carol_2.south) edge[->,>=stealth, bend right=15, dashed] node[below] {$\textsf{Pr}(c_4|u)$} (Alice_2.south);

\end{tikzpicture}
\end{adjustbox}
\begin{adjustbox}{max width=\linewidth}
\begin{tikzpicture}
\tikzstyle{arrow} = [ultra thick,->,>=stealth]
\tikzstyle{line} = [thick]
\tikzstyle{member} = [fill=tblue!20]
\tikzstyle{skill} = [fill=torange!20]
\tikzstyle{title} = [fill=tgreen!20]
\tikzstyle{company} = [fill=tpurple!20]

\newcommand{\myoffset}{-3.5}

\node at (-5.2, 0) {$p \rightarrow \mathcal{F}(p, \mathcal{G})$:};

\node[draw, member, name=Alice_1] at (0+\myoffset,0) {Alice};
\node[draw, member, name=Bob_1] at (1.6+\myoffset,0) {Bob};
\node[draw, member, name=Dan_1] at (3.2+\myoffset,0) {Dan};

\draw [arrow] (Alice_1) -- (Bob_1);
\draw [arrow] (Bob_1) -- (Dan_1);

\draw [arrow, color=tred] (4.+\myoffset, 0) -- node[color=black, align=center]{node-structure\\expansion} (6.+\myoffset, 0);

\node[draw, member, name=Alice_2] at (7+\myoffset, 0) {Alice};
\node[draw, member, name=Bob_2] at (9.5+\myoffset, 0) {Bob};
\node[draw, member, name=Dan_2] at (12+\myoffset, 0) {Dan};

\draw [arrow] (Alice_2) -- (Bob_2);
\draw [arrow] (Bob_2) -- (Dan_2);

\node[draw, skill, name=sklearn] at (7+\myoffset, .7) {sklearn};
\node[draw, title, name=ml_engineer] at (7+\myoffset, -.7) {ML Engineer};

\draw[line] (Alice_2) -- (sklearn);
\draw[line] (Alice_2) -- (ml_engineer);

\node[draw, skill, name=rl] at (9.5+\myoffset, .7) {Rep. Learning};
\node[draw, company, name=lkdn] at (9.5+\myoffset, -.7) {LinkedIn};

\draw[line] (Bob_2) -- (rl);
\draw[line] (Bob_2) -- (lkdn);

\node[draw, title, name=sr] at (12+\myoffset, .7) {Sr. ML Engineer};
\node[draw, skill, name=dl] at (12+\myoffset, -.7) {Deep Learning};

\draw[line] (Dan_2) -- (dl);
\draw[line] (Dan_2) -- (sr);

\draw [arrow, color=tpurple, dashed] (13+\myoffset, 0) -- node[color=black, align=center]{structure-based\\skip-gram} (15+\myoffset, 0);

\node[draw, member, name=Alice_2, opacity=.4, dashed] at (16+\myoffset, 0) {Alice};
\node[draw, member, name=Bob_2, ultra thick] at (18.5+\myoffset, 0) {Bob};
\node[draw, member, name=Dan_2] at (21+\myoffset, 0) {Dan};

\draw [arrow, opacity=.4] (Alice_2) -- (Bob_2);
\draw [arrow, opacity=.4] (Bob_2) -- (Dan_2);

\node[draw, skill, name=sklearn, opacity=.4, dashed] at (16+\myoffset, .7) {sklearn};
\node[draw, title, name=ml_engineer] at (16+\myoffset, -.7) {ML Engineer};

\draw[line] (Alice_2) -- (sklearn);
\draw[line] (Alice_2) -- (ml_engineer);

\node[draw, skill, name=rl] at (18.5+\myoffset, .7) {Rep. Learning};
\node[draw, company, name=lkdn] at (18.5+\myoffset, -.7) {LinkedIn};

\draw[line] (Bob_2) -- (rl);
\draw[line] (Bob_2) -- (lkdn);

\node[draw, title, name=sr, opacity=.4, dashed] at (21+\myoffset, .7) {Sr. ML Engineer};
\node[draw, skill, name=dl, opacity=.4, dashed] at (21+\myoffset, -.7) {Deep Learning};

\draw[line] (Dan_2) -- (dl);
\draw[line] (Dan_2) -- (sr);

\path[draw]
    (Bob_2.west) edge [->,>=stealth, dashed, bend left=90] node[above left] {$\textsf{Pr}(c_1|u)$} (rl.west)
    (Bob_2.east) edge [->,>=stealth, dashed, bend left=90] node[right] {$\textsf{Pr}(c_2|u)$} (lkdn.east)
    (Bob_2.west) edge [->,>=stealth, dashed, bend left=55] node[above] {$\textsf{Pr}(c_3|u)$} (ml_engineer.east)
    (Bob_2.east) edge [->,>=stealth, dashed, bend left=55] node [below] {$\textsf{Pr}(c_4|u)$} (Dan_2);

\end{tikzpicture}
\end{adjustbox}

\caption{Path generation and embedding learning example of Star2Vec. Star2Vec first generates a length-$2$ person path using $\mathcal{P}$, then expands the path into stars of size $k_s = 2$ using $\mathcal{F}$, and finally optimizes the network representation of \texttt{Bob} using a structure-based skip-gram with window size $k_w=5$. Comparing to traditional path-based skip-gram models show at the top of the figure, Star2Vec uses social connection strength-aware random walk and therefore can walk on social connections with high strength (\texttt{Alice}-\texttt{Bob}) and unobserved but highly similar persons (\texttt{Bob}-\texttt{Dan}). Star2Vec is also likely to optimize nodes with semantically similar person and entity neighbors because of its node-structure expansion function.}
\label{fig:expansion}
\end{figure*}
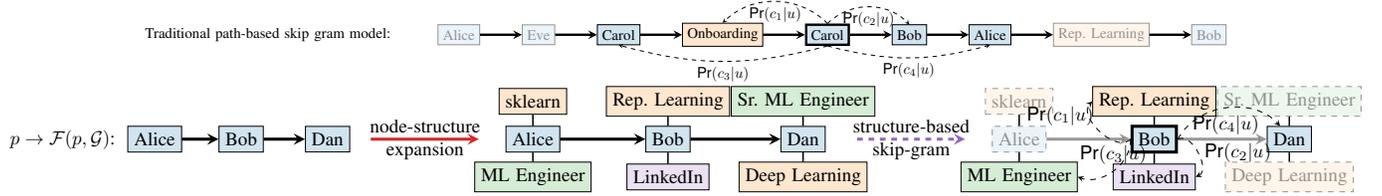

\hl{In this section, we present Star2Vec and its details in three parts: (A) its social connection strength-aware, random walk-based path generation method, (B) its node-structure expansion-based path augmentation, and (C) its star-structure-based person and entity representation learning method.}

\subsection{\hl{Social Connection Strength-aware Biased Random Walk}}

Models that operate on heterogeneous networks typically enumerate constrained network paths so that the nodes on the path conform to a sequence of types~\cite{lao2010relational}. The path constraints are typically called metapaths, and they are hand-curated by a human designer. Specific metapaths are meaningful for specific tasks; a typical example found in the related literature~\cite{sun2013pathselclus} suggests that the path \textit{author}$\rightarrow$\textit{paper}$\leftarrow$\textit{author}, which represents co-authorship in a bibliographical heterogeneous network, is important to identify communities of researchers. \hl{However, it is not always clear which metapaths are meaningful for heterogeneous professional social networks with many node types, and human curators may miss important information or introduce bias into the model. Moreover, the social connection ambiguity nature of HPSNs and many social networks makes manually composing reliable metapath even harder.} In fact our results in Sec.~\ref{sec:experiments} show that \hl{on networks with ambiguous social connections, such as LinkedIn and Facebook}, representations learned from ``intuitive'' metapaths are even worse than the ones learned from homogeneous models. 


\hl{As we discussed above, social connections in HPSNs are oftentimes ambiguous and have different connection strengths, which makes it almost impossible to design reliable metapaths for representation learning. To better model such ambiguous social connections in HPSNs and general social networks, one natural approach is to label each social connection with its true relationship type. However, this labelling task is problematic for two reasons. First, without sufficient signals beyond the connection itself, the interpretation of a social connection can be subjective. Second, the social connection between people evolves over time.}

\hl{Luckily, if an HPSN model can treat each social connection differently based on its connection strength, then we no longer need to disambiguate each person-to-person edge manually. Moreover, by modeling the connection strength between two persons, the model can even discover and leverage unobserved social connections. Recall that the result of ambiguous social connections is that different person-to-person edges with different social strengths are grouped into the same edge type and treated equally. If we can properly model the connection strength based on the network context, then we can weight social connections using their connection strengths and reduce the importance of social connections such as neighbors and friends which represent ambiguous relationships.}


\hl{To estimate the strength of social connections, we need to first identify useful network context that can help model the social connection. According to Def.~\ref{def:HSN}, we separate the nodes in HPSNs into two groups, person-type nodes and non-person-type entity nodes.
Although person nodes are the same and person-to-person connections are often ambiguous across different social networks, non-person entity types and person-to-entity connections usually represent the special interests of each social network and therefore are reliable and have consistent semantic meanings within the same relationship type. For example, unambiguous person-to-entity relationships on Facebook include person-to-political preference and person-likes-post, whereas on LinkedIn, unambiguous professional person-to-entity relationships such as person-knows-skill, person-has-title, person-worksAt-company, etc.}

Based on this observation, we assume that edges between person and non-person entity nodes \hl{(also called attribute nodes) are often unambiguous, because they represent the characteristics of the person nodes. With this assumption, next we will discuss how to estimate the strength of social connections using other unambiguous person-to-entity relationships.}

Consider the example illustrated in Fig.~\ref{fig:graph}, where the \hl{social connection \texttt{Alice}-to-\texttt{Bob} is an ambiguous person-to-person connection and its reliability is hard to determine by simply examining a single edge. By looking at other alternative, unambiguous person-to-entity connections among \texttt{Alice} and \texttt{Bob}, for example person $\xrightarrow{worksAt}$ company and \textit{person} $\xrightarrow{knows}$ \textit{skill}, we know \texttt{Alice}-\texttt{Bob} is a stronger social connection than \texttt{Alice}-\texttt{Carol} because \texttt{Alice} and \texttt{Bob} are more structurally similar because they are both indirectly connected via many unambiguous person-to-entity paths. Hence, we can estimate the connection strength of such person-to-person edges by modeling unambiguous alternative person-to-entity paths between two nodes. Here we formally define the social connection strength as follows} 

\begin{mydef}\label{def:reliable_connection}
\hl{Given an HPSN $\mathcal{G}$ and a social connection edge $u\xrightarrow{r}v$ between two person nodes $u$ and $v$, the \textbf{social connection strength} of $u\xrightarrow{r}v$ is defined by some support function $\mathcal{S}(u, r, v, \mathcal{G})$ that measures the structural similarity between $u$ and $v$ with respect to some social connection relationship $r$.}
\end{mydef}

To model the structural similarity between nodes using alternative unambiguous connections between two person with respect to some social relationship $r$, \hl{we borrow the concept of second-order proximity from LINE~\cite{tang2015line} and define the support function $\mathcal{S}$ of some social connection $u \xrightarrow{r} v$ as their heterogeneous second-order proximity.}

\vspace{-.2cm}
\begin{small}
\begin{equation}\label{eq:ap_score}
\mathcal{S}(u, r, v, \mathcal{G})=\frac{\sum_{x\in N(u, \mathbf{D}_{r})} \frac{\mathbb{I}(x,v)}{\textrm{deg}(x, \phi(v))} }{|N(u, \mathbf{D}_{r})|},
\end{equation}
\end{small}
\vspace{-.2cm}

\noindent{}where $r\in\mathbf{R}_{ppl}$, the relationship dependent neighbor set $N(u,\mathbf{D}_r)=\{x|\phi(x)\in\mathbf{D}_r, (u,x)\in\mathbf{E}\}$ represents the neighbors $x$ of person $u$ with node type $\phi(x)\in\mathbf{D_r}$, $\mathbb{I}(x,v)$ is an indicator function testing $(x, v)\in\mathbf{E}$, and $\textrm{deg}(x,t)$ returns the number of type-$t$ nodes that $x$ connects to. One can also view Eq.~\ref{eq:ap_score} as a function measuring the heterogeneous second-order proximity based on the probability that $u$ can reach $v$ in two steps using only nodes within a given dependency type set $\mathbf{D}_{r}$. Note that one can also use Eq.~\ref{eq:ap_score} to measure the connection strength of some unobserved social connection $u\xrightarrow{r}v$. When $\mathbf{D}_r = \mathbf{T}$ holds for all $r\in\mathbf{R}$, then Eq.~\ref{eq:ap_score} degenerates to a homogeneous higher-order proximity scoring function~\cite{tang2015line}.

To generate dependency set $\mathbf{D}_r$ \hl{with respect to social connection relationship $r$, one could first collect some $u\xrightarrow{r}v$ examples, or use association rule mining~\cite{galarraga2013amie}, or predicate path mining~\cite{shi2016discriminative} to discover associated length-$2$ paths, and then extract all intermediate non-person node types to construct $\mathbf{D}_r$. For example, in LinkedIn network, the dependency-set of a simple person-to-person connection is $\mathbf{D}_{connect} = \{\textit{title},\textit{skill}\}$, which defines the social connection strength of mutual connections by their common titles and skills. $\mathbf{D}_r$ can also be manually defined,~\eg, the dependency set of LinkedIn's follower-influencer social relationship $\mathbf{D}_{follow}=\{\textit{member},\textit{industry}\}$ if we are interested in modeling the connection strength based on influencer popularity among their followers' social circle and common industry experience.}


In Eq.~\ref{eq:ap_score} we limit the order of the proximity to $2$ to simplify the computation, but it can be easily extended to higher-orders by modeling paths instead of neighboring nodes. 

With Eq.~\ref{eq:ap_score}, we define the epsilon-greedy style \hl{social connection strength}-aware transition function $\mathcal{P}$ that determines the transition score from \hl{person $u_i$ to person $u_{i+1}$} as

\vspace{-.2cm}
\begin{small}
\begin{equation}\label{eq:walk2}
\resizebox{.9\hsize}{!}{$\displaystyle
\begin{split}
    \mathcal{P}(u_i, u_{i+1}, \mathcal{G}) & = \underbrace{(1-\alpha)\sum\limits_{r_k \in \mathbf{R}_{(u_i, u_{i+1})}} \mathcal{S}(u_i, r_k, u_{i+1}, \mathcal{G})}_{\text{walk on existing social connections}}\\
    & \underbrace{+\  \alpha\sum\limits_{r\in\mathbf{R}_{ppl}}S(u_i,r,u_{i+1},\mathcal{G})}_{\text{walk on unobserved social connections}}
\end{split}
$}
\end{equation}
\end{small}
\vspace{-.2cm}

\noindent{}\hl{where the relationship set $R_{(u_i, u_{i+1})}$ contains all relationships $r_k$ that connect $u_i$ and $u_{i+1}$ in $\mathcal{G}$, $r$ is some person-to-person relationship type from $\mathbf{R}_{ppl}$, which is defined in the HPSN $\mathcal{G}$, and $\alpha$ is some jump probability that allows the model to walk on unobserved but highly possible social connections between highly similar nodes measured by $\mathcal{S}$. To reduce the computational complexity on enumerating all possible $u_{i+1}$, we limit $u_{i+1}$ to person-type nodes that can be reached from person $u_i$ within two steps.}

Equation~\ref{eq:walk2} addresses the problem of ambiguous social connections by calculating the \hl{social connection strength} of person-to-person edges to avoid walking over \hl{ambiguous social connections that do not contribute to the professional similarity}, such as \texttt{Alice}$\rightarrow$\texttt{Carol} in Fig.~\ref{fig:graph}. Moreover, by considering the transition score between highly similar but not directly connected persons \hl{(second term in Eq.~\ref{eq:walk2})}, \hl{the model will also generate social connection paths that are not directly observed.} For example, \texttt{Alice}$\rightarrow$\texttt{Bob} could be extended to \texttt{Alice}$\rightarrow$\texttt{Bob}$\rightarrow$\texttt{Dan} using $\mathcal{P}$ even though two structurally similar nodes \texttt{Bob} and \texttt{Dan} are not directly connected.



Note that the random walker used in previous works~\cite{perozzi2014deepwalk,hamilton2017representation} is a special case of Eq.~\ref{eq:walk2}, where $\alpha=0$, $u_* \in\mathbf{V}$, $|\mathbf{R}|=1$, and $\mathcal{S}(u, r, v, \mathcal{G})=\frac{1}{N(u)}$.

\subsection{Node-structure Expansion}

In the previous section we described a \hl{social connection strength}-aware random walk that generates paths using \hl{observed and unobserved social connections with high connection strengths defined by Eq.~\ref{eq:ap_score}}. However, \hl{due to the lack of non-person entities in the generated paths, this method cannot learn person and entity embeddings jointly. If we explicitly generate additional paths by specifying  certain metapaths,~\eg, person-skill-person or person-title-person}, then the representations of each entity type is likely to be trained disjointedly, which would produce incomparable node representations across different node types. \hl{Such an approach may yield good results in a node-type clustering visualization, but cannot be used for cross-type inference, such as suggesting skills to members, finding related skills given a title or find important companies at some location, etc.}

To remedy this issue and learn person and entity embeddings in the same semantic space so different type of entities can be compared directly, we apply an extra node-structure expansion post-processing function $\mathcal{F}$ on the path set $\mathbf{P}$ to generate an expanded, diversified person-entity structure (\ie, a star) path set $\mathbf{P}^\prime$ to increase the entity-type coverage in $\mathbf{P}$ and appropriately capture higher-order heterogeneous proximity in the model.


To describe $\mathcal{F}$, first recall that network representation learning models inspired by Word2Vec view the nodes and paths as words and sentences respectively, and learn node representations by maximizing the similarity between a node in some path and its surrounding nodes. 

In Star2Vec, we extend this idea by replacing the single person $u$ in the path with a star-structure $s(u)$ containing $k_s$ neighbors of $u$ with $u$ as the star's center. To continue the analogy, $s(u)$ essentially becomes a ``phrase'' in the overall sentence, and we use nodes in nearby phrases to update the representation of nodes in $s(u)$. By doing so, we can increase the context node similarity and diversity within a given window size comparing to other models. In Fig.~\ref{fig:expansion} we illustrate this node expansion using a length-$2$ path generated by $\mathcal{P}$ and expanding each node $u$ in the path to $s(u)$ with a star size $k_s=2$. Here we define $\mathcal{F}$ as $\mathcal{F}(p, \mathcal{G}) \rightarrow \{s(u_1)\rightarrow\cdots\rightarrow s(u_l) | u_i \in p \}$, $s(u)=\{u\}\cup\{v_i | v_i \sim \textsf{Pr}(v_i|u,\mathcal{G}), v_i \in N(u,\mathbf{T})\}$ , $|s(u)|=k_s + 1$, and $\textsf{Pr}(v|u,\pi, \mathcal{G})$ is some disproportionate stratified sampling probability defined as

\vspace{-.1cm}
\begin{small}
\begin{equation}\label{eq:expansion_prob}
\textsf{Pr}(v|u,\pi, \mathcal{G}) \propto \frac{\pi^{\phi(v)}}{|N(u, \{\phi(v)\})|},
\end{equation}
\end{small}
\vspace{-.1cm}

\noindent{}where $\pi\in\mathbb{R}^{|\mathbf{T}|}$ is the parameter of $t \sim \mathsf{Multi}(t|\mathbf{\pi})$, $\pi^{\phi(v)}$ denotes the probability of selecting node type $\phi(v)$. $\pi$ can be approximated by the confidence score of $\phi(u_{i})\leadsto\phi(u_{i+1})$ via $T_i$ for $T_i \in \mathbf{T}$ using AMIE~\cite{galarraga2013amie} or some simple distributions such as uniform distribution. $|N(u, \{\phi(v)\})|$ is the number of $\phi(v)$ typed-nodes that connect to $u$.

\subsection{Structure-based Skip-gram}

\hl{After we construct the star-structured paths $\mathbf{P}^\prime$, next we discuss how to learn person and entity embeddings using $\mathbf{P}^\prime$.} In a standard random walk-based network representation learning setting, the objective is often defined as 

\begin{small}
\begin{equation}
    \argmax_\theta \sum_{u \in \mathbf{V}} \sum_{c\in C(u)} \log(\textsf{Pr}(c|u;\theta)),
\end{equation}
\end{small}

\noindent{}where $c$ is the context node of $u$ defined by $C(u)$, which is usually the neighbor of $u$ in a random walk path $p$ within a window size $k_w$. Here we can not apply this objective directly to the proposed Star2Vec model because it is unclear about how to generate context nodes for $u$ from star-shaped structure paths $s(u_1)\leadsto s(u_l)$ instead of simple node paths $u_1\leadsto u_l$. 

So, we first define $k_w^\prime = \left\lceil k_w/k_s \right\rceil$ that represents the smallest star window size that covers at least $k_w$ nodes. The context nodes of $u$ within a star window of size $k_w^\prime$ on $s(u_1)\leadsto s(u_l)$ is then defined as

\vspace{-.2cm}
\begin{small}
\begin{equation}\label{eq:context_superset}
V_{c}^{u, p} = \bigcup_{j = i - \frac{k^\prime_w}{2}}^{i+\frac{k^\prime_w}{2}}s(u_{j}) \setminus \{u\}, u \in s(u_i),
\end{equation}
\end{small}
\vspace{-.2cm}

\noindent{}in which $s(u_i)$ is the star-shaped structure centered at $u_i$ and $V_{c}^{u,p}$ is the superset of the context nodes of $u$ in path $p$. We then extract $k_w$ context nodes for $u\in s(u_i)$ by randomly sampling from $V_{c}^{u,p}$ and rewrite the objective as

\vspace{-.2cm}
\begin{small}
\begin{equation}\label{eq:stucture-skip-gram}
\argmax_\theta \sum_{u \in V} \sum_{p\in\mathbf{P}^\prime}\sum_{i}^{k_w}\mathbb{E}_{c_i \sim \mathsf{Unif}(V_c^{u,p})} \log(\textsf{Pr}(c_i|u;\theta)),
\end{equation}
\vspace{-.2cm}
\end{small}

\noindent{}where the conditional probability $\textsf{Pr}(c|u;\theta)$ is actually the log-normalized score of the embedding inner product defined as $\exp(\mathbf{W}_c\cdot\mathbf{W}_u^T) / \sum_{v\in \mathbf{V}} \exp(\mathbf{W}_v\cdot\mathbf{W}_u^T)$, in which $\mathbf{W}\in\mathbb{R}^{|\mathbf{V}|\times d}$ is the embedding matrix. We combine a negative sampling function with Eq.~\ref{eq:stucture-skip-gram} to define the loss function

\vspace{-.2cm}
\begin{small}
\begin{equation}\label{eq:loss}
\begin{split}
    \mathcal{L} =& \sum_{u \in \mathbf{V}} \sum_{p\in\mathbf{P}^\prime}\sum_{i}^{k_w}\mathbb{E}_{c_i \sim \mathsf{Unif}(V_c^{u,p})}\bigg(\log(\sigma(\mathbf{W}_{c_i}\cdot\mathbf{W}_u^T))\\
    &+ \sum_{j}^{k_\textsf{neg}}\mathbb{E}_{v_j\sim\mathsf{Dist}(u)}\log(\sigma(-\mathbf{W}_{v_j}\cdot\mathbf{W}_u^T))\bigg),
\end{split}
\end{equation}
\end{small}
\vspace{-.2cm}

\noindent{}in which $\sigma(x)=1/(1+\exp(-x))$, $\textrm{Dist}(u)$ is some negative sampling distribution, $k_\textrm{neg}$ and $k_w$ are the number of negative samples and context window size respectively. We follow the convention of previous works~\cite{mikolov2013efficient} and use stochastic gradient descent with back propagation to optimize Eq.~\ref{eq:loss}. 

Training time complexity of Star2Vec is $O(V)$ which is the same as homogeneous network embedding models~\cite{perozzi2014deepwalk,grover2016node2vec} and lower than rich feature models' $O(V^2)$~\cite{huang2017label}.

\section{Experiments}\label{sec:experiments}

\hl{We compare Star2Vec to other representation learning models on LinkedIn's Economic Graph and a public available subset of Facebook using a variety of tasks including industry and social circle classification, skill / title clustering, and member-entity link predictions.}
We also conduct extensive case studies to demonstrate the possibilities of using the representations learned from Star2Vec to solve real-world tasks LinkedIn is facing such as skill recommendation, career suggestion, and general entity retrieval on the LinkedIn's Economic Graph.

\begin{table}[t]
    \caption{Evaluation datasets used in this work.}
    \label{tab:dataset}
    \centering
    \begin{adjustbox}{max width=.9\linewidth}
    \begin{tabular}{r rrrr}
    Dataset       & \#Nodes & \#Members & \#NodeType & \#Edges \\
    \toprule
    Facebook        & $6,319$ & $4,039$ & $29$ & $127,777$ \\
    LinkedIn-60k  & $\sim60$K & $\sim14$K & $10$ & $>500$K \\
    LinkedIn-44M  & $\sim44$M & $\sim40$M & $10$ & $>6$B \\
    \end{tabular}
    \end{adjustbox}
\end{table}

\begin{table*}[ht]
    \caption{Person-to-industry classification on LinkedIn-60k.}
    \label{tab:classification_lkdn60k}
    \centering
    \begin{adjustbox}{max width=.8\linewidth}
        \begin{tabular}{c l c c c c c c c c c c}
        & &\multicolumn{10}{c}{Training Set \%} \\\cline{3-12}
        Metric & Model & $5\%$ & $10\%$ & $20\%$ & $30\%$ & $40\%$ & $50\%$ & $60\%$ & $70\%$ & $80\%$ & $90\%$ \\\toprule
        \multirow{5}{*}{Macro-F1} & Metapath2Vec & $0.0733$ & $0.1060$ & $0.1291$ & $0.1491$ & $0.1587$ & $0.1658$ & $0.1747$ & $0.1746$ & $0.1794$ & $0.1811$\\
        & LINE (1+2) & $0.0191$ & $0.0199$ & $0.0218$ & $0.0234$ & $0.0248$ & $0.0261$ & $0.0270$ & $0.0282$ & $0.0286$ & $0.0302$\\
        & DeepWalk & $\mathbf{0.1105}$ & $\mathbf{0.1381}$ & $0.1558$ & $0.1724$ & $0.1825$ & $0.1890$ & $0.1916$ & $0.1963$ & $0.1977$ & $0.1945$ \\
        & Node2Vec & $0.0504$ & $0.0888$ & $0.1251$ & $0.1457$ & $0.1648$ & $0.1759$ & $0.1864$ & $0.1954$ & $0.1981$ & $0.1961$ \\\cline{2-12}
        & TransE   & $0.0240$ & $0.0274$ & $0.0358$ & $0.0436$ & $0.0491$ & $0.0537$ & $0.0556$ & $0.0582$ & $0.0592$ & $0.0596$ \\
        & TransR   & $0.0245$ & $0.0281$ & $0.0375$ & $0.0461$ & $0.0521$ & $0.0567$ & $0.0599$ & $0.0615$ & $0.0649$ & $0.0640$ \\
        & ProjE    & $0.0677$ & $0.0889$ & $0.1065$ & $0.1225$ & $0.1300$ & $0.1373$ & $0.1429$ & $0.1472$ & $0.1503$ & $0.1415$ \\\cline{2-12}
        & \textbf{Star2Vec} & $0.0943$ & $0.1334$ & $\mathbf{0.1617}$ & $\mathbf{0.1802}$ & $\mathbf{0.1902}$ & $\mathbf{0.1979}$ & $\mathbf{0.2056}$ & $\mathbf{0.2113}$ & $\mathbf{0.2131}$ & $\mathbf{0.2149}$ \\\midrule
        \multirow{5}{*}{Micro-F1} & Metapath2Vec &  $0.4888$ & $0.5076$ & $0.5215$ & $0.5305$ & $0.5336$ & $0.5355$ & $0.5390$ & $0.5387$ & $0.5421$ & $0.5411$\\
        & LINE (1+2) & $0.4422$ & $0.4477$ & $0.4515$ & $0.4543$ & $0.4578$ & $0.4603$ & $0.4618$ & $0.4654$ & $0.4664$ & $0.4675$ \\
        & DeepWalk & $0.4990$ & $0.5158$ & $0.5258$ & $0.5302$ & $0.5364$ & $0.5373$ & $0.5373$ & $0.5391$ & $0.5434$ & $0.5432$\\
        & Node2Vec & $0.4621$ & $0.4919$ & $0.5140$ & $0.5246$ & $0.5316$ & $0.5330$ & $0.5357$ & $0.5399$ & $0.5434$ & $0.5402$\\\cline{2-12}
        & TransE   & $0.4567$ & $0.4648$ & $0.4773$ & $0.4845$ & $0.4862$ & $0.4902$ & $0.4914$ & $0.4937$ & $0.4961$ & $0.4949$ \\
        & TransR   & $0.4619$ & $0.4700$ & $0.4834$ & $0.4903$ & $0.4940$ & $0.4964$ & $0.4991$ & $0.4999$ & $0.5033$ & $0.5068$ \\
        & ProjE    & $0.4108$ & $0.4339$ & $0.4629$ & $0.4811$ & $0.4882$ & $0.4971$ & $0.5028$ & $0.5077$ & $0.5104$ & $0.5118$ \\\cline{2-12}
        & \textbf{Star2Vec} & $\mathbf{0.5038}$ & $\mathbf{0.5197}$ & $\mathbf{0.5317}$ & $\mathbf{0.5368}$ & $\mathbf{0.5410}$ & $\mathbf{0.5436}$ & $\mathbf{0.5434}$ & $\mathbf{0.5466}$ & $\mathbf{0.5505}$ & $\mathbf{0.5500}$\\
        \end{tabular}
    \end{adjustbox}
\end{table*}
\begin{table*}[ht]
    \caption{Person-to-social circle classification on Facebook.}
    \label{tab:classification_fb}
    \centering
    \begin{adjustbox}{max width=.8\linewidth}
        \begin{tabular}{c l c c c c c c c c c c}

        & & \multicolumn{10}{c}{Training Set \%} \\\cline{3-12}
        Metric & Model & $5\%$ & $10\%$ & $20\%$ & $30\%$ & $40\%$ & $50\%$ & $60\%$ & $70\%$ & $80\%$ & $90\%$ \\\toprule
        \multirow{5}{*}{Macro-F1} & Metapath2Vec & $0.0504$ & $0.0958$ & $0.1338$ & $0.1690$ & $0.1960$ & $0.2167$ & $0.2414$ & $0.2603$ & $0.2666$ & $0.2426$\\
        & LINE (1+2) & $0.1240$ & $0.1697$ & $0.2338$ & $0.2616$ & $0.2951$ & $0.3143$ & $0.3302$ & $0.3509$ & $0.3507$ & $0.3371$\\
        & DeepWalk & $0.2510$ & $0.3249$ & $\mathbf{0.4106}$ & $0.4509$ & $0.4875$ & $0.5048$ & $0.5160$ & $0.5250$ & $0.5137$ & $0.5019$\\
        & Node2Vec & $0.0517$ & $0.1129$ & $0.1949$ & $0.2395$ & $0.2703$ & $0.2833$ & $0.3053$ & $0.3190$ & $0.3292$ & $0.3351$\\\cline{2-12}
        & TransE   & $0.0755$ & $0.1175$ & $0.1589$ & $0.1960$ & $0.2117$ & $0.2305$ & $0.2443$ & $0.2618$ & $0.2746$ & $0.2830$ \\
        & TransR   & $0.0778$ & $0.1197$ & $0.1625$ & $0.1982$ & $0.2178$ & $0.2393$ & $0.2586$ & $0.2772$ & $0.2848$ & $0.2885$ \\
        & ProjE    & $\mathbf{0.3102}$ & $\mathbf{0.3336}$ & $0.3620$ & $0.3669$ & $0.3663$ & $0.3659$ & $0.3598$ & $0.3595$ & $0.3603$ & $0.3507$ \\\cline{2-12}
        & \textbf{Star2Vec} & $0.1495$ & $0.2390$ & $0.3943$ & $\mathbf{0.5055}$ & $\mathbf{0.5687}$ & $\mathbf{0.6210}$ & $\mathbf{0.6588}$ & $\mathbf{0.6866}$ & $\mathbf{0.6977}$ & $\mathbf{0.6135}$\\\midrule
        \multirow{5}{*}{Micro-F1} & Metapath2Vec &  $0.2233$ & $0.3346$ & $0.4139$ & $0.4434$ & $0.4746$ & $0.4899$ & $0.5050$ & $0.5251$ & $0.5296$ & $0.5305$\\
        & LINE (1+2) & $0.4683$ & $0.5292$ & $0.6038$ & $0.6240$ & $0.6499$ & $0.6579$ & $0.6693$ & $0.6934$ & $0.6990$ & $0.6958$\\
        & DeepWalk & $\mathbf{0.6319}$ & $\mathbf{0.6947}$ & $\mathbf{0.7422}$ & $0.7659$ & $0.7811$ & $0.7899$ & $0.7984$ & $0.7994$ & $0.8000$ & $0.8070$ \\
        & Node2Vec & $0.2490$ & $0.4593$ & $0.6146$ & $0.6660$ & $0.6962$ & $0.7051$ & $0.7207$ & $0.7268$ & $0.7392$ & $0.7411$\\\cline{2-12}
        & TransE   & $0.4176$ & $0.4705$ & $0.5144$ & $0.5456$ & $0.5548$ & $0.5655$ & $0.5680$ & $0.5727$ & $0.5791$ & $0.5834$ \\
        & TransR   & $0.4198$ & $0.4714$ & $0.5163$ & $0.5454$ & $0.5564$ & $0.5668$ & $0.5723$ & $0.5776$ & $0.5818$ & $0.5845$ \\
        & ProjE    & $0.5290$ & $0.5389$ & $0.5421$ & $0.5363$ & $0.5333$ & $0.5281$ & $0.5282$ & $0.5215$ & $0.5290$ & $0.5243$ \\\cline{2-12}
        & \textbf{Star2Vec} & $0.4435$ & $0.5903$ & $0.7145$ & $\mathbf{0.7769}$ & $\mathbf{0.8101}$ & $\mathbf{0.8307}$ & $\mathbf{0.8511}$ & $\mathbf{0.8583}$ & $\mathbf{0.8661}$ & $\mathbf{0.8760}$\\
        \end{tabular}
    \end{adjustbox}
\end{table*}
\begin{table*}[ht]
    \caption{Person-to-industry classification on LinkedIn-44M.}
    \label{tab:classification_lkdn44m}
    \centering
    \begin{adjustbox}{max width=.8\linewidth}
        \begin{tabular}{c l c c c c c c c c c c}
        & & \multicolumn{10}{c}{Training Set \%} \\\cline{3-12}
        Metric & Model & $5\%$ & $10\%$ & $20\%$ & $30\%$ & $40\%$ & $50\%$ & $60\%$ & $70\%$ & $80\%$ & $90\%$ \\\toprule
        \multirow{4}{*}{Macro-F1} & Metapath2Vec & $0.1688$ & $0.1912$ & $0.2097$ & $0.2195$ & $0.2246$ & $0.2285$ & $0.2315$ & $0.2325$ & $0.2343$ & $0.2354$ \\
        & DeepWalk & $0.1820$ & $0.2069$ & $0.2250$ & $0.2347$ & $0.2410$ & $0.2467$ & $0.2481$ & $0.2489$ & $0.2526$ & $0.2517$ \\
        & Node2Vec & $0.1943$ & $0.2218$ & $0.2437$ & $0.2539$ & $0.2606$ & $0.2658$ & $0.2683$ & $0.2691$ & $0.2707$ & $0.2692$ \\\cline{2-12}
        & \textbf{Star2Vec} & $\mathbf{0.2421}$ & $\mathbf{0.2685}$ & $\mathbf{0.2908}$ & $\mathbf{0.3012}$ & $\mathbf{0.3067}$ & $\mathbf{0.3108}$ & $\mathbf{0.3122}$ & $\mathbf{0.3153}$ & $\mathbf{0.3170}$ & $\mathbf{0.3172}$ \\\midrule
        \multirow{4}{*}{Micro-F1} & Metapath2Vec & $0.4101$ & $0.4280$ & $0.4419$ & $0.4473$ & $0.4502$ & $0.4516$ & $0.4540$ & $0.4548$ & $0.4564$ & $0.4558$ \\
        & DeepWalk & $0.4113$ & $0.4284$ & $0.4394$ & $0.4445$ & $0.4476$ & $0.4501$ & $0.4508$ & $0.4514$ & $0.4519$ & $0.4508$ \\
        & Node2Vec & $0.4259$ & $0.4434$ & $0.4555$ & $0.4607$ & $0.4637$ & $0.4664$ & $0.4671$ & $0.4678$ & $0.4694$ & $0.4674$ \\\cline{2-12}
        & \textbf{Star2Vec} & $\mathbf{0.4783}$ & $\mathbf{0.4919}$ & $\mathbf{0.5019}$ & $\mathbf{0.5069}$ & $\mathbf{0.5089}$ & $\mathbf{0.5101}$ & $\mathbf{0.5114}$ & $\mathbf{0.5122}$ & $\mathbf{0.5140}$ & $\mathbf{0.5151}$ \\
        \end{tabular}
    \end{adjustbox}
    \vspace{-.1cm}
\end{table*}

\subsection{Datasets}
\hl{We consider LinkedIn as a heterogeneous professional social network and Facebook as a general-purpose heterogeneous social networks with different focus on the social connections and entity types}. Table~\ref{tab:dataset} shows a summary of the three datasets. LinkedIn-60k and LinkedIn-44M are two subsets of LinkedIn's Economic Graph and Facebook is derived from Facebook-egonet~\cite{leskovec2012learning}.



\subsection{Experiment Setup}

We compared Star2Vec with node embedding methods and knowledge graph completion methods. Other matrix factorization methods as well as rich-feature methods could not be compared because of their high time complexity or the lack of certain features. We use the best performing parameters reported in each work for all tasks on the Facebook and LinkedIn-60k. On the LinkedIn-44M network we limited the embedding size to $64$ and only generate $10$ length-$100$ paths per entity node for all node embedding models in order to manage the memory and disk consumption. We generated metapaths for the Metapath2Vec model by enumerating all length-$2$ person to entity metapaths as suggested in the original work. As in prior work, all networks are treated as undirected graphs to avoid creating random walk sinks. 

\hl{On the LinkedIn datasets, we set the dependency set of person-to-person connection as $D_{connectTo} = \{$ \textit{title}, \textit{skill}, \textit{company}, \textit{school} $\}$. On the Facebook dataset, $D_{connectTo}$ contains all non-person entity types. We set $\pi^{\phi(\star)}$ to be a uniform distribution for all datasets. As for other hyper-parameters, we conduct a hyper-parameter test for the proposed Star2Vec model and report the results at the end of this section.} We were unable to gather results of LINE and knowledge graph completion models on LinkedIn-44M due to scaling issues. The model was trained on a single machine with $48$ cores, and it took $9$ hours to converge on the LinkedIn-44M dataset.

\subsection{Multi-class and Multi-label Classification}

First we explore the effectiveness of Star2Vec on multi-class and multi-label classifications. In both cases we use external labels with at least $10$ members and train logistic regression models on top of the learned embeddings to perform the prediction. We vary the training size from $5\%$ to $90\%$ using a stratified split w.r.t each class and treat the remaining data as testing set. We repeat each experiment setting $10$ times and report the average Macro-F1 and Micro-F1 scores.


We perform the multi-class classification by predicting a \textit{person}s' self-reported \textit{industry} on LinkedIn because each person has exact one label. Similarly, we perform multi-label classification task on Facebook where social circles are used as labels. Note that users may belong to multiple social circles so a single user can have multiple labels.


The classification results of LinkedIn-60k and LinkedIn-44M are shown in Tab.~\ref{tab:classification_lkdn60k} and Tab.~\ref{tab:classification_lkdn44m}. Due to memory limitation and high time complexity, we are not able to compare LINE and Knowledge Graph Completion based models on the LinkedIn-44M dataset. On both networks Star2Vec outperformed other models on Macro-F1 by up to $37.8\%$ \ignore{up to $10.2\%$ (relatively $37.8\%$)} and up to $10.2\%$\ignore{ $4.7\%$ (relatively $10.2\%$)} on Micro-F1. Interestingly, the improvement was more significant on the large-scale LinkedIn-44M network. We believe the difference in improvement is because the smaller LinkedIn-60k network is well-curated so that the network is more complete and the number of ambiguous social connections is limited. The large improvement on the less-curated LinkedIn-44M network indicates that Star2Vec is robust on networks with ambiguous social connections \hl{due to its social connection strength-aware random walk}.

The results of the multi-label classification task on the Facebook network are shown in Tab.~\ref{tab:classification_fb}. We find that Star2Vec works well especially when the amount of training data is greater than 30\%. This improvement indicates that Star2Vec is able to \hl{learn person embeddings that better capture a person's characteristics by modeling non-person entities with people jointly using the node-structure expansion function.}


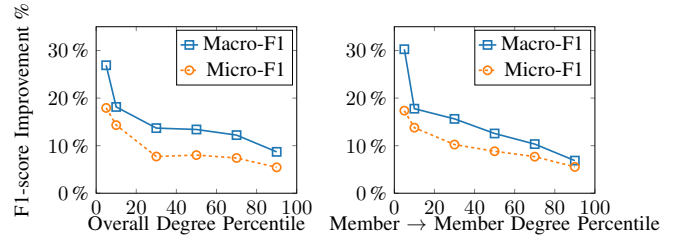
\begin{figure}[t]
    \centering
        \begin{adjustbox}{max width=\linewidth}
            \newcommand{\stratylabel}{F1-score Improvement \%}
  
\begin{tikzpicture}
\begin{axis}[
    name=deg,
    xlabel={Overall Degree Percentile},
    y label style={at={(axis description cs:-0.1,.5)},anchor=south},
    ylabel={\stratylabel},
    xmin=0,xmax=100,
    ymin=0,ymax=35,
    yticklabel=\pgfmathprintnumber{\tick}\,\%,
    legend cell align=right,
    legend style={nodes={scale=2, transform shape}},
    every axis plot/.append style={ultra thick},
    every tick label/.append style={scale=2},
    label style={scale=2}
]

\addplot[color=tblue, mark=square, mark size=4pt]
    table[x=percentile, y=increase] {plots/data/stratified_by_degree_macro.txt};
\addplot[color=torange, mark=o, dashed, mark options={solid}, mark size=4pt]
    table[x=percentile, y=increase] {plots/data/stratified_by_degree_micro.txt};
\legend{Macro-F1, Micro-F1};
\end{axis}
\end{tikzpicture}

\begin{tikzpicture}
\begin{axis}[
    name=member,
    xlabel={Member $\rightarrow$ Member Degree Percentile},
    xmin=0,xmax=100,
    ymin=0,ymax=35,
    yticklabel=\pgfmathprintnumber{\tick}\,\%,
    legend cell align=right,
    legend style={nodes={scale=2, transform shape}},
    every axis plot/.append style={ultra thick},
    every tick label/.append style={scale=2},
    label style={scale=2}
]

\addplot[color=tblue, mark=square, mark size=4pt]
    table[x=percentile, y=increase] {plots/data/stratified_by_member_macro.txt};
\addplot[color=torange, mark=o, dashed, mark options={solid}, mark size=4pt]
    table[x=percentile, y=increase] {plots/data/stratified_by_member_micro.txt};
\legend{Macro-F1, Micro-F1};
\end{axis}
\end{tikzpicture}
        \end{adjustbox}
        \caption{Classification improvement stratified by degree percentile \hl{(in ascending order)} on LinkedIn-44M. Percentage is based on second best model.}
        \label{fig:stratified_member_degree}
        \vspace{-.5cm}
\end{figure}

\begin{table*}[t]
    \caption{Top-$3$ similarity results of provided queries on LinkedIn-44M. }
    \label{tab:case_study}
    \centering
    \newcounter{magicrownumbers}
    \newcommand\rownumber{\stepcounter{magicrownumbers}\arabic{magicrownumbers}}
    \begin{adjustbox}{max width=\linewidth}
        \begin{tabular}{l l | l l l l l}
        ~  & (Query, target node type) & \multicolumn{3}{c@{\quad}}{Top-$3$ Results} \\\toprule
        \rownumber& (Software Dev, \textit{title}) & Junior Software Engineer & Software Dev Team Lead & Software Dev Contractor \\
        \rownumber& (Software Dev $+$ KDB, \textit{title}) & Quantitative Dev	& Financial Software Dev & Front Office Dev \\
        \rownumber& (Sr. Audit Accountant, \textit{title}) & Supervising Sr. Accountant & Audit Sr. & Audit Staff Accountant \\
        \rownumber& (Sr. Accountant Audit $-$ Member of AICPA, \textit{title}) & SVP Marketing Business Development & SVP Strategy Business Development & SVP Human Resources Administration \\
        \rownumber& (FBI, \textit{skill}) & Counterintelligence & Federal Law Enforcement & Cybercrime Investigation \\
        \rownumber & (Medical Research, \textit{region}) & Iowa City, Iowa Area & Washington D.C. Metro Area & Gainesville, Florida Area \\
        \rownumber& (Deloitte, Title) & Sr. Advisory Consultant & Sr. Manager Advisory Services & Sr. Associate Advisory \\
        \rownumber & (Deep Learning, \textit{skill}) & Machine Learning & Artificial Neural Networks & Neural Networks \\
        \end{tabular}
    \end{adjustbox}
    \vspace{-.4cm}
\end{table*}

\hl{Because many social networks are incomplete and follow a power-law degree distribution, the ability of handling poorly connected nodes with many unobserved social connections is a crucial factor for an HPSN representation learning model. To better understand Star2Vec's ability to learn embeddings for nodes with limited connectivity}, we further group the \textit{person}-nodes in LinkedIn-44M by their degree and plot the F1 improvement over the second-best performing model stratified by the degree percentile in Fig.~\ref{fig:stratified_member_degree}. These results clearly demonstrates that Star2Vec works better than the best performing model especially on nodes with low connectivity (left-hand side of plots). \hl{This indicates Star2Vec's social connection-strength based path generation can implicitly increase the connectivity of nodes with few connections, and therefore learn better representations compared to other methods.}


\subsection{Node Clustering}

Next, we performed three node clustering tasks on the LinkedIn-44M network. We compare models on the Adjusted Mutual Information (AMI)~\cite{vinh2010information}. For these clustering tasks, we used the trained node embeddings as input and assigned nodes to clusters using $k$-means. The cluster labels used in each task are summarized as follows: 1) in the \textit{skill}-to-\textit{domain} task we assigned each \textit{skill} to a single hand-curated domain,~\eg, \texttt{Java} and \texttt{Python} are assigned to \texttt{Computing}; 2) the \textit{person}-to-\textit{industry} task uses the same label as described in the classification task; and 3) the \textit{title}-to-\textit{specialty} task divides job titles by induced specialties,~\eg, \texttt{Junior Java Dev} would belong to \texttt{Software Developer}. In each task, the number of clusters $k$ is set to equal the number of labels observed in the network. We present the results in Tab.~\ref{tab:node_clustering}.

\begin{table}[t]
    \caption{Adjusted Mutual Information (AMI) score of node clustering results on LinkedIn-44M.}
    \label{tab:node_clustering}
    \centering
    \begin{adjustbox}{max width=.9\linewidth}
    \begin{tabular}{r c c c}
        ~ & \textit{skill}-to-\textit{domain} & \textit{person}-to-\textit{industry} & \textit{title}-to-\textit{specialty} \\
        \toprule
        Metapath2Vec & $0.5965$ & $0.1105$ & $0.3878$ \\
        DeepWalk & $0.5918$ & $0.1689$ & $0.3663$ \\
        Node2Vec & $0.5969$ & $0.1633$ & $0.3850$ \\\hline
        \textbf{Star2Vec} & $\mathbf{0.5992}$ & $\mathbf{0.3294}$ & $\mathbf{0.4200}$ 
    \end{tabular}
    \end{adjustbox}
    \vspace{-.2cm}
\end{table}

We find that Star2Vec performs well on these tasks. However, the performance boost is most pronounced the person-to-industry task. This performance boost is due to differences in the connectivity patterns between person and non-person entity nodes. Recall that in HPSNs ambiguous connections are mostly person-to-person social connections, which means person nodes are more vulnerable to ambiguity because, in prior models, their representations were mainly defined by their peers. On the other hand, entity nodes such as skill and title are less likely to be affected because they usually have a robust second-order connectivity pattern, \ie, a skill and a title are likely to be similar if they connect to the same person.

This also explains why Metapath2Vec works well on some clustering tasks, but not others. Under these conditions, metapaths such as person-entity-person tend to decouple entity node types from each other during training. This decoupling leads to poor person embeddings and poor clustering performance. 

\subsection{Link Prediction}

\begin{table}[t]
    \caption{AUROC score of link prediction on LinkedIn-44M.}
    \label{tab:link_prediction}
    \centering
    \begin{adjustbox}{max width=.9\linewidth}
        \begin{tabular}{r c c c c}
        ~ & \textit{person}-\textit{region} & \textit{person}-\textit{industry} & \textit{person}-\textit{skill} \\\toprule
        Metapath2Vec & $0.5097$ & $0.5189$ & $0.5356$ \\
        DeepWalk & $0.5034$ & $0.5034$ & $0.5489$ \\
        Node2Vec & $0.5013$ & $0.5080$ & $0.5605$ \\\hline
        \textbf{Star2Vec} & $\mathbf{0.6370}$ & $\mathbf{0.5840}$ & $\mathbf{0.7175}$ \\

        \end{tabular}
    \end{adjustbox}
    \vspace{-.2cm}
\end{table}

In addition to the classification and clustering tasks, we also evaluate Star2Vec on multiple person-entity link prediction tasks. Here we use $70\%$ of the data for training, $5\%$ for validation, and evaluate the model performance on the remaining $25\%$ with the same amount of negative edges, which is generated by replacing a node on an existing edge with an incorrect random node of the same type. This strategy results in a more difficult but more realistic link prediction problem comparing to previous settings~\cite{grover2016node2vec}. To speed up the evaluation on the LinkedIn-44M dataset, we sampled $50,000$ edges at random from $1$ billion testing edges.

The area under the ROC (AUROC) score of link prediction tasks are shown in Tab.~\ref{tab:link_prediction}. These consistent performance improvements indicate the proposed model can learn embeddings that better captures the semantic meaning of nodes.

\begin{figure*}[t]
    \centering
    \begin{adjustbox}{max width=.99\linewidth}
        \begin{adjustbox}{max width=.48\linewidth}
            \pgfplotsset{
    MyParaSentStyle/.style={
        legend cell align=left,
        every axis plot/.append style={ultra thick},
        label style={scale=1.5},
        legend style={
        nodes={scale=1.5, transform shape},
        legend image post style={scale=1.5},
        },
        every axis plot post/.append style={
            every mark/.append style={scale=2}
        }
    }
}

\begin{tikzpicture}
\begin{axis}[
    MyParaSentStyle,
    xlabel={window size $k_w$},
    ylabel={Classification Task (F1)},
    xmin=5,xmax=13,
    ymin=.1,ymax=.8,
]

\addplot[color=tblue, mark=x]
    table[x=para, y=Macro.F1] {./plots/data/window_size.txt};
\addplot[color=torange, mark=o, dashed, mark options={solid}]
    table[x=para, y=Micro.F1] {./plots/data/window_size.txt};
\legend{Macro-F1, Micro-F1};

\end{axis}
\end{tikzpicture}

\begin{tikzpicture}
\begin{axis}[
    MyParaSentStyle,
    xlabel={embedding size $d$},
    xmin=16,xmax=256,
    ymin=.1,ymax=.8,
]

\addplot[color=tblue, mark=x]
    table[x=para, y=Macro.F1] {./plots/data/embed_size.txt};
\addplot[color=torange, mark=o, dashed, mark options={solid}]
    table[x=para, y=Micro.F1] {./plots/data/embed_size.txt};
\legend{Macro-F1, Micro-F1};

\end{axis}
\end{tikzpicture}

\begin{tikzpicture}
\begin{axis}[
    MyParaSentStyle,
    xlabel={star size $k_s$},
    xmin=1,xmax=10,
    ymin=.1,ymax=.8,
]

\addplot[color=tblue, mark=x]
    table[x=para, y=Macro.F1] {./plots/data/star_size.txt};
\addplot[color=torange, mark=o, dashed, mark options={solid}]
    table[x=para, y=Micro.F1] {./plots/data/star_size.txt};
\legend{Macro-F1, Micro-F1};

\end{axis}
\end{tikzpicture} 
        \end{adjustbox}
        \begin{adjustbox}{max width=.48\linewidth}
            \begin{tikzpicture}
\begin{axis}[
    xlabel={walk length $l$},
    ylabel={Classification Task (F1)},
    xmin=25,xmax=175,
    ymin=.1,ymax=.8,
    legend cell align=left,
    every axis plot/.append style={ultra thick},
    label style={scale=1.5},
    legend style={
        nodes={scale=1.5, transform shape},
        legend image post style={scale=1.5},
        },
    every axis plot post/.append style={
        every mark/.append style={scale=2}
    }
]

\addplot[color=tblue, mark=x]
    table[x=para, y=Macro.F1] {./plots/data/walk_length.txt};
\addplot[color=torange, mark=o, dashed, mark options={solid}]
    table[x=para, y=Micro.F1] {./plots/data/walk_length.txt};
\legend{Macro-F1, Micro-F1};

\end{axis}
\end{tikzpicture}

\begin{tikzpicture}
\begin{axis}[
    xlabel={walk per node $w$},
    xmin=1,xmax=20,
    ymin=.1,ymax=.8,
    legend cell align=left,
    every axis plot/.append style={ultra thick},
    label style={scale=1.5},
    legend style={
        nodes={scale=1.5, transform shape},
        legend image post style={scale=1.5},
        },
    every axis plot post/.append style={
        every mark/.append style={scale=2}
    }
]

\addplot[color=tblue, mark=x]
    table[x=para, y=Macro.F1] {./plots/data/walk_per_node.txt};
\addplot[color=torange, mark=o, dashed, mark options={solid}]
    table[x=para, y=Micro.F1] {./plots/data/walk_per_node.txt};
\legend{Macro-F1, Micro-F1};

\end{axis}
\end{tikzpicture}

\begin{tikzpicture}
\begin{axis}[
    xlabel={jump probability $\alpha$ (\%)},
    xmin=0,xmax=95,
    ymin=.1,ymax=.8,
    legend cell align=left,
    every axis plot/.append style={ultra thick},
    label style={scale=1.5},
    legend style={
        nodes={scale=1.5, transform shape},
        legend image post style={scale=1.5},
        },
    every axis plot post/.append style={
        every mark/.append style={scale=2}
    }
]

\addplot[color=tblue, mark=x]
    table[x=para, y=Macro.F1] {./plots/data/rej_prob.txt};
\addplot[color=torange, mark=o, dashed, mark options={solid}]
    table[x=para, y=Micro.F1] {./plots/data/rej_prob.txt};
\legend{Macro-F1, Micro-F1};

\end{axis}
\end{tikzpicture}
        \end{adjustbox}
    \end{adjustbox}
    \begin{adjustbox}{max width=.99\linewidth}
        \begin{adjustbox}{max width=.48\linewidth}
            \begin{tikzpicture}
\begin{axis}[
    xlabel={window size $k_w$},
    ylabel={Clustering Task (AMI)},
    xmin=5,xmax=13,
    ymin=0.2,ymax=1,
    legend cell align=left,
    every axis plot/.append style={ultra thick},
    every tick label/.append style={scale=1.5},
    label style={scale=1.5},
    legend style={
        nodes={scale=1.5, transform shape},
        legend image post style={scale=1.5},
        },
    every axis plot post/.append style={
        every mark/.append style={scale=2}
    }
]

\addplot[color=tblue, mark=x]
    table[x=para, y=skill_domain] {./plots/data/window_size.txt};
\addplot[color=torange, mark=o, dashed, mark options={solid}]
    table[x=para, y=member_industry] {./plots/data/window_size.txt}; 
\addplot[color=tgreen, mark=diamond*, dashed, mark options={solid}]
    table[x=para, y=title_specialty] {./plots/data/window_size.txt};
\legend{Skill-Domain, Member-Industry, Title-Specialty};

\end{axis}
\end{tikzpicture}

\begin{tikzpicture}
\begin{axis}[
    xlabel={embedding size $d$},
    xmin=16,xmax=256,
    ymin=0.2,ymax=1,
    legend cell align=left,
    every axis plot/.append style={ultra thick},
    every tick label/.append style={scale=1.5},
    label style={scale=1.5},
    legend style={
        nodes={scale=1.5, transform shape},
        legend image post style={scale=1.5},
        },
    every axis plot post/.append style={
        every mark/.append style={scale=2}
    }
]

\addplot[color=tblue, mark=x]
    table[x=para, y=skill_domain] {./plots/data/embed_size.txt};
\addplot[color=torange, mark=o, dashed, mark options={solid}]
    table[x=para, y=member_industry] {./plots/data/embed_size.txt}; 
\addplot[color=tgreen, mark=diamond*, dashed, mark options={solid}]
    table[x=para, y=title_specialty] {./plots/data/embed_size.txt};
\legend{Skill-Domain, Member-Industry, Title-Specialty};

\end{axis}
\end{tikzpicture}

\begin{tikzpicture}
\begin{axis}[
    xlabel={star size $k_s$},
    xmin=1,xmax=10,
    ymin=0.2,ymax=1,
    legend cell align=left,
    every axis plot/.append style={ultra thick},
    every tick label/.append style={scale=1.5},
    label style={scale=1.5},
    legend style={
        nodes={scale=1.5, transform shape},
        legend image post style={scale=1.5},
        },
    every axis plot post/.append style={
        every mark/.append style={scale=2}
    }
]

\addplot[color=tblue, mark=x]
    table[x=para, y=skill_domain] {./plots/data/star_size.txt};
\addplot[color=torange, mark=o, dashed, mark options={solid}]
    table[x=para, y=member_industry] {./plots/data/star_size.txt}; 
\addplot[color=tgreen, mark=diamond*, dashed, mark options={solid}]
    table[x=para, y=title_specialty] {./plots/data/star_size.txt};
\legend{Skill-Domain, Member-Industry, Title-Specialty};

\end{axis}
\end{tikzpicture} 
        \end{adjustbox}
        \begin{adjustbox}{max width=.48\linewidth}
            \begin{tikzpicture}
\begin{axis}[
    xlabel={walk length $l$},
    ylabel={Clustering Task (AMI)},
    xmin=25,xmax=175,
    ymin=0.2,ymax=1,
    legend cell align=left,
    every axis plot/.append style={ultra thick},
    every tick label/.append style={scale=1.5},
    label style={scale=1.5},
    legend style={
        nodes={scale=1.5, transform shape},
        legend image post style={scale=1.5},
        },
    every axis plot post/.append style={
        every mark/.append style={scale=2}
    }
]

\addplot[color=tblue, mark=x]
    table[x=para, y=skill_domain] {./plots/data/walk_length.txt};
\addplot[color=torange, mark=o, dashed, mark options={solid}]
    table[x=para, y=member_industry] {./plots/data/walk_length.txt}; 
\addplot[color=tgreen, mark=diamond*, dashed, mark options={solid}]
    table[x=para, y=title_specialty] {./plots/data/walk_length.txt};
\legend{Skill-Domain, Member-Industry, Title-Specialty};

\end{axis}
\end{tikzpicture}

\begin{tikzpicture}
\begin{axis}[
    xlabel={walk per node $w$},
    xmin=1,xmax=20,
    ymin=0.2,ymax=1,
    legend cell align=left,
    every axis plot/.append style={ultra thick},
    every tick label/.append style={scale=1.5},
    label style={scale=1.5},
    legend style={
        nodes={scale=1.5, transform shape},
        legend image post style={scale=1.5},
        },
    every axis plot post/.append style={
        every mark/.append style={scale=2}
    }
]

\addplot[color=tblue, mark=x]
    table[x=para, y=skill_domain] {./plots/data/walk_per_node.txt};
\addplot[color=torange, mark=o, dashed, mark options={solid}]
    table[x=para, y=member_industry] {./plots/data/walk_per_node.txt}; 
\addplot[color=tgreen, mark=diamond*, dashed, mark options={solid}]
    table[x=para, y=title_specialty] {./plots/data/walk_per_node.txt};
\legend{Skill-Domain, Member-Industry, Title-Specialty};

\end{axis}
\end{tikzpicture}

\begin{tikzpicture}
\begin{axis}[
    xlabel={jump probability $\alpha$ (\%)},
    xmin=0,xmax=95,
    ymin=0.2,ymax=1,
    legend cell align=left,
    every axis plot/.append style={ultra thick},
    every tick label/.append style={scale=1.5},
    label style={scale=1.5},
    legend style={
        nodes={scale=1.5, transform shape},
        legend image post style={scale=1.5},
        },
    every axis plot post/.append style={
        every mark/.append style={scale=2}
    }
]

\addplot[color=tblue, mark=x]
    table[x=para, y=skill_domain] {./plots/data/rej_prob.txt};
\addplot[color=torange, mark=o, dashed, mark options={solid}]
    table[x=para, y=member_industry] {./plots/data/rej_prob.txt}; 
\addplot[color=tgreen, mark=diamond*, dashed, mark options={solid}]
    table[x=para, y=title_specialty] {./plots/data/rej_prob.txt};
\legend{Skill-Domain, Member-Industry, Title-Specialty};

\end{axis}
\end{tikzpicture}
        \end{adjustbox}
    \end{adjustbox}
    \vspace{-.4cm}
    \caption{Parameter sensitivity test for classification and clustering tasks on LinkedIn-44M.}
    \label{fig:classification_parameter_sensitivity}
\end{figure*}
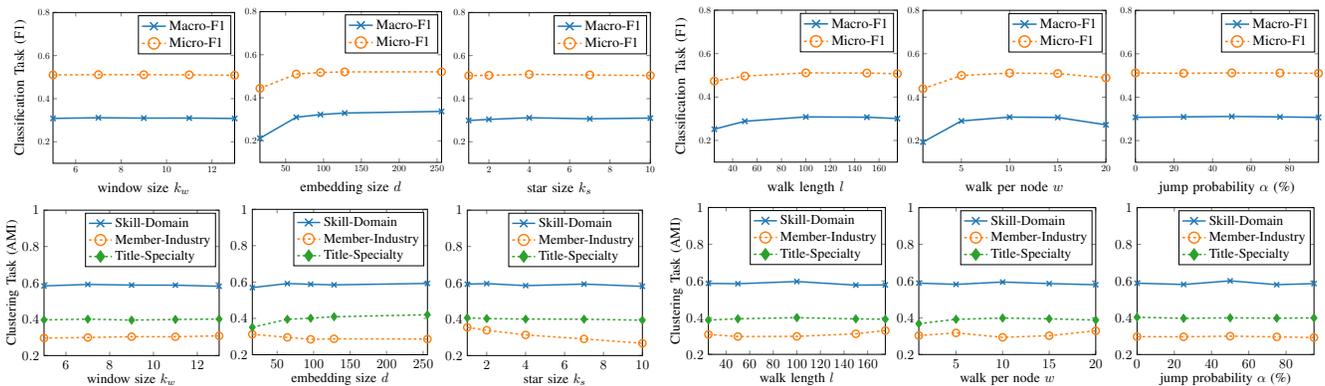

\subsection{Case Studies}

In addition to the \hl{applications that can be formed into} standard network analysis tasks, we are also interested in applying the learned representations to other practical cases and to gain insights from the HPSN. In this section, we demonstrate how to employ the learned Star2Vec embeddings to solve interesting \hl{entity retrieval} tasks. We built a nearest neighbor model that takes, as input, a query vector and a target node type, and returns the top-$k$ nearest nodes (in terms of cosine similarity). The top-$3$ results of $8$ example queries are shown in Tab.~\ref{tab:case_study}. Note that there are no direct relationships between any of these query objects.


\vspace{.2cm}
\noindent{\textbf{Next Career Move}}. The \textit{title} of a \textit{person} typically depends on their \textit{skill} set and experience. Thus, learning new skills could potentially lead to new opportunities. To capture such a change, we combine the vector of a \textit{person}'s current \textit{title} with their most recent \textit{skill} and suggest \textit{title}s based on the combined representation~\cite{li2017nemo}. In Tab.~\ref{tab:case_study}, we see that after a \texttt{software developer} learns a new \textit{skill} \texttt{KDB}, which is a financial database, the job recommendation shifts from general \texttt{software development} to jobs in the financial sector. 

\vspace{.2cm}
\noindent{\textbf{Alternative Career Suggestion}}. Oftentimes, individuals may be interested in a particular job \textit{title}, but do not possess the necessary \textit{skill} set. In this case, we may wish to show the alternative job \textit{title}s that do not require certain \textit{skill}s. We achieve this goal by subtracting the missing \textit{skill} from the \textit{title}'s representation. Tab.~\ref{tab:case_study} shows alternative job recommendations for \texttt{senior audit accountant} without requiring a CPA license. Note that the returned alternative jobs also preserves the seniority of the given job title.

\vspace{.2cm}
\noindent{\textbf{General Similarity Search}}. The learned representations can also perform general similarity searches between arbitrary nodes regardless of node types. These searches can be used to infer relationships that are absent from the graph. These inferred relationships allow members to gain valuable insights to better their place in the network.

\subsection{Hyper-parameter Sensitivity}

Here we study the hyper-parameter sensitivity of Star2Vec and report the relationship between Star2Vec performance as a function of its various hyper-parameters. These include the window size $k_w$, the node embedding size $d$, the star size $k_s$, the random walk length $l$, the number of walks per node $w$, and the jump probability $\alpha$. The hyper-parameter sensitivity results on two tasks are illustrated in Fig.~\ref{fig:classification_parameter_sensitivity}. We find that the performance of Star2Vec is stable and largely insensitive to most of the hyper-parameters. The primary exception is, as excepted, the embedding size $d$. Another particularly interesting finding is that the model's performance remains largely unchanged when $k_s$ or $k_w$ is changed. This demonstrates that training with long, reliable simple paths can achieve results similar to models trained with shorter, star-structured paths covering the same number of nodes.

\section{Related Work}\label{sec:related_work}
A variety of Representation Learning (RL) models have been developed to learn network embeddings in recent years. Here we group them by their inputs.

\vspace{.2cm}
\noindent{}\textbf{Homogeneous Network Embedding}. Inspired by Word2Vec~\cite{mikolov2013distributed}, a number of random walk-based RL models have been proposed~\cite{perozzi2014deepwalk,grover2016node2vec,nguyen2018dynamic}. Just as Word2Vec updates each word embedding to match those within the same sentence, these models update each node-embedding to match its neighboring nodes on truncated random walk paths. LINE~\cite{tang2015line} and HOPE~\cite{ou2016asymmetric} take a different approach and learn node-embeddings through matrix factorization-like objectives. SNDE~\cite{wang2016structural}, on the other hand, uses an auto-encoder to learn node-embeddings from second-order proximity data. MVE~\cite{qu2017multiview} separates a network into multiple views and learn node representations using attention. \hl{CTDNE~\cite{nguyen2018dynamic} uses a temporal random walk method to consider the temporal neighborhood.} Despite their differences, these models only use un-typed topological information and therefore miss the rich information encoded in the node types.  

\vspace{.2cm}
\noindent{\textbf{Network Embedding with Rich Features}.} To address the limitations that accompany homogeneous network embedding, several models have been proposed to augment the network in order to generate better network representations. LANE~\cite{huang2017label} uses an auxiliary attribute network and node labels to learn node embeddings jointly. SNE~\cite{liao2017sne}, on the other hand, encode attributes into embeddings and utilizes MLP to learn node representations. TriDNR~\cite{pan2016tri} uses two skip-gram models to jointly train node embeddings from a node's content and neighbors. Although these methods have shown promising results by augmenting the graph with additional features, they either have a high $O(V^2)$ time complexity or require additional features which may not be available in all networks. Moreover, these models were evaluated on limited sized networks, which is not applicable to real-world online networks, such as Economic Graph's million-node scale. Recently, RGCN~\cite{zhu_robust_2019} studies learning robust network embeddings against small deliberate perturbations in graph structures, but it does not address how to handle existing ambiguous social connections that appear in many real-world social networks.

\vspace{.2cm}
\noindent{\textbf{Heterogeneous Network Embedding}.} Metapath2Vec~\cite{dong2017metapath2vec} is a graph embedding model that guides the random walk with human-curated metapaths over HINs. Other models have extended this idea into meta-graph walks~\cite{fionda2017meta,jiang2017semi}. However, the problem of generating informative metapaths is still unclear. \hl{ImVerde designed a vertex-diminished random walk method to boost the probability of visiting nodes from the same class~\cite{wu2018imverde}. This approach can better separate nodes from different classes but will penalize inter-class relatedness}. Knowledge Graph Completion methods~\cite{bordes2013,lin2015learning,shi2017proje,dettmers2017convolutional, shi2018open} can be viewed as heterogeneous embedding models, but they are explicitly designed for link prediction and are not well suited for node classification and clustering tasks, especially when the relationship type is absent.


\section{Conclusions and Future Work}\label{sec:conclusion}
In this work, we presented a heterogeneous professional social network representation learning model that 1) \hl{addresses the ambiguous social connection and incomplete network problem by designing a social connection strength-aware random walk method without introducing additional model parameters, and 2) utilizes rich entity types to learn person and entity embeddings jointly with a node-structure expansion function.} Star2Vec outperforms existing models on three heterogeneous social network datasets across different scales and tasks, which highlights the necessity to rectify \hl{ambiguous social connections in heterogeneous social networks, leveraging unobserved social connections, and modeling person and entity embeddings jointly}. We also conducted extensive case studies to demonstrate how to use these embeddings to discover professional insights and power other recommendation tasks.

As for future work, we will incorporate rich contextual features into Star2Vec in a scalable way. Another interesting extension would be to further improve the model's ability to estimate the social connection strength.

\vspace{.2cm}
\noindent\textbf{Acknowledgements.} 
This work is partially supported by the US Army Research Office (ARO W911NF-17-1-0448).

\bibliographystyle{IEEEtran}
\bibliography{acmart}

\end{document}